\documentclass[aps,prd,preprint,superscriptaddress,showpacs,floatfix,nobibnotes]{revtex4-1}

\usepackage{latexsym}
\usepackage{amsmath}
\usepackage{amssymb}
\usepackage{graphicx}

\usepackage{bm}
\usepackage{epsfig}
\usepackage{subfigure}
\newcommand{\bea}{\begin{eqnarray}}
\newcommand{\eea}{\end{eqnarray}}

\begin{document}

\title{Dynamical gap generation in graphene with frequency dependent renormalization effects}

\author{M.E. Carrington}
\email[]{carrington@brandonu.ca} \affiliation{Department of Physics, Brandon University, Brandon, Manitoba, R7A 6A9 Canada}\affiliation{Winnipeg Institute for Theoretical Physics, Winnipeg, Manitoba}

\author{C.S. Fischer}
\email[]{} \affiliation{Institut f\"{u}r Theoretische Physik, Justus-Liebig-Universit\"{a}t Giessen, Heinrich-Buff-Ring 16, 35392 Giessen, Germany}

\author{L. von Smekal}
\email[]{} \affiliation{Institut f\"{u}r Theoretische Physik, Justus-Liebig-Universit\"{a}t Giessen, Heinrich-Buff-Ring 16, 35392 Giessen, Germany}

\author{M.H. Thoma}
\email[]{} \affiliation{I. Physikalisches Institut, Justus-Liebig-Universit\"{a}t Giessen, Heinrich-Buff-Ring 16, 35392 Giessen, Germany}

\date{\today}

\begin{abstract}
  We study the frequency dependencies in the renormalization of the fermion Greens function for the $\pi$-band electrons in graphene and their influence on the dynamical gap generation at sufficiently strong interaction. Adopting the effective QED-like description for the low-energy excitations within the Dirac-cone region we self consistently solve the fermion Dyson-Schwinger equation in various approximations for the photon propagator and the vertex function with special emphasis on frequency dependent Lindhard screening and retardation effects. 
\end{abstract}


\normalsize
\maketitle

\normalsize

\section{Introduction}
\label{introduction-section}

In recent years there has been much interest in the study of graphene (two reviews are Refs.~\cite{neto1,neto2}).
Graphene is a 2-dimensional crystal of carbon atoms which exhibits many unique electronic properties and some interesting quantum effects. 
Some of the interest in graphene is due to its possible applications in a wide range of technological fields. 
Theoretically graphene is interesting to physicists, in part, because it provides a  condensed matter analogue of many problems that are studied in particle physics using relativistic quantum field theory, including topological phase transitions, chiral symmetry breaking, and strong coupling dynamics.

We consider the simplest form, mono-layer graphene, in which the carbon atoms are arranged in a 2-dimensional hexagonal lattice. 
Due to this particular lattice structure, the low energy dynamics are described by a continuum quantum field theory
in which the electronic quasi-particles have a linear Dirac-like dispersion relation of the form $E=\pm v_F p$ where 
$v_F\sim c/300$ is the velocity of a massless electron in graphene. The Dirac-like parts of the Brillouin zone are called the Dirac cones, and the apex of the cones where the quasi-particle energies go to zero are called the Dirac points. 
We will consider the theory at half filling (zero chemical potential), in which case 
the band structure is such that the quasi-particle density of states vanishes when the energies of the quasi-particle excitations go to zero.
The consequence is that electrons in graphene cannot screen in the normal metallic sense, and thus even in the absence of a gap the system is not a true metal, but is referred to as a semi-metal.

An important question is whether or not the quasi-particle interactions are strong enough to produce a gap and cause the system to undergo a phase transition to an insulating state. 
From a technological point of view, a finite gap would make graphene more promising as a potential material for producing novel electronic devices. 
A gap would also be theoretically interesting as a concrete realization of the phenomenon of chiral symmetry breaking \cite{gusynin07} which has been studied in particle physics for many years. 
The generation of this gap would correspond to the dynamical breaking of the symmetry under interchange of the two triangular sub-lattices which make up the hexagon.

Measurements of the conductivity of suspended graphene have shown that the effective coupling is not strong enough to produce a gap, and that the insulating state is therefore not physically realizable \cite{elias11}. 
However, the experimental observation of fairly strong Fermi velocity renormalization effects might indicate that it is not too far from a Mott insulating state, so that a transition could still be induced e.g. by mechanical strain or by an external magnetic field via magnetic catalysis \cite{brane}.

%
The effective coupling has the form
$\alpha = \frac{e^2}{4\pi \epsilon \hbar v_F}$ where 
$\epsilon$ is related to the physical properties of the graphene sheet.
In vacuum (suspended graphene) $\epsilon=1$, but if the graphene is immersed in another material or attached to a substrate it would have  higher value $\epsilon>1$.
Physically this means the medium or substrate generates dielectric screening that lowers the effective fine structure constant of the system. The maximum possible effective coupling is therefore obtained with the vacuum value $\epsilon=1$ and is about $\alpha_{\rm max} = 2.2$. 
In order to determine theoretically if a gap is formed in the physical system, one calculates the critical coupling for gap formation. If it is larger than 2.2, we conclude that the physical interactions are not strong enough to produce an insulating phase. Note that we are necessarily working with a strongly coupled system, which means that perturbative methods are not applicable. 

Lattice simulations can provide reliable results for strongly coupled condensed matter systems, if the bare interactions correctly describe the physical system. However, even in a theory where the microscopic theory is completely specified (like QCD), continuum, infinite volume, and chiral extrapolations are often difficult. 
Non-perturbative continuum approaches like functional renormalization group, Dyson-Schwinger equations, and $n$-particle irreducible theories provide valuable alternative approaches, although each of these methods has its own shortcomings, notably the necessity to introduce some kind of truncation. 

Many theoretical calculations of the critical coupling for gap formation have been done \cite{khv,brane,liu,gor,case,gonz,drut,hands,pop,sols,pop,rong-liu}. Early studies based on effective low energy theories for the Dirac cone region typically all obtained critical couplings around $\alpha_c \sim 1$, and hence less than $\alpha_{\rm max}$ in contradiciton with experiment. It has now been widely accepted that this contradiction is resolved by using realistically screened Coulomb interactions, especially at short distances, such as those obtained from a constrained random phase approximation in Ref.~\cite{wehling11}. Such interactions have been used in ab-initio Hybrid-Monte-Carlo simulations on a hexagonal lattice to show that the resulting critical coupling is indeed larger than $ \alpha_{\rm max}$ \cite{ulybyshev13,lorenz1}.

In this paper, we are especially interested in the influences of
frequency dependent Lindhard screening and retardation effects. In particular, we assess the resulting frequency dependences in the fermionic Greens function which have usually been neglected in Dyson-Schwinger studies. We do not necessarily aim at a realistic desription of graphene which would have to include the screening from the $\sigma$-band electrons and localized higher energy states as in Refs.~\cite{wehling11,ulybyshev13,lorenz1}. Instead we adopt the effective QED-like description for the low-energy excitations within the Dirac-cone region. Within this description, however, we systematically investigate the various frequency dependencies including retardation effects beyond the non-relativistic Coulomb interaction. Additional screening of the short distance part of the interaction in more realistic calculations will generally tend to increase all critical couplings that we obtain in the present study. One general consequence of frequency dependent renormalization effects is that they also require vertex corrections. It then becomes of practical importance to devise truncations beyond the bare vertex approximation. As a result of our study we can identify a comparatively simple vertex truncation as a very good compromise between computing efforts and the full construction based on gauge invariance of Ref.~\cite{ball-chu}. In fact, this simple truncation provides a much better approximation here than one might have expected from analogous studies in QED in 2+1 dimensions \cite{fischer-2004}.

Our goal is to generalize the calculation for graphene as much as possible and determine which of the assumptions that have been used in the past have a significant effect on the result. In particular, several different approximations have been introduced to simplify different frequency integrals, i.e. the zeroth component of the momentum integral (see equation (\ref{d4K})). We discuss the physical motivation for these approximations and study numerically the effect of relaxing them. The model we are using is described below and the calculation is described in detail in Section \ref{notation-section}. 

The Euclidean action of the low energy effective theory is given by
\bea
\label{action}
S=\int d^3 x \sum_{a}\bar\psi_a \left(i\partial_\mu -e A_\mu\right)M_{\mu\nu}\gamma_\nu \psi _a - \frac{\epsilon}{4e^2}\int d^3x F_{\mu\nu}\frac{1}{2\sqrt{-\partial^2}}F_{\mu\nu} \text{ + gauge fixing}
\eea
where the Greek indices take values $\in\{0,1,2\}$. 
The fermionic part of the action looks like that of a free Dirac theory with a linear dispersion relation. This reflects the fact that the low energy effective theory is a valid description of the system close to the Dirac points. 
Four-component Dirac spinors are used for quasi-particle excitations on
both sub-lattices, with momenta close to either of the two Dirac points.
The true spin of the electrons formally appears as an additional flavor
quantum number, and we take $N_f = 2$ for monolayer graphene. The three 4-dimensional
$\gamma$-matrices form a reducible representation of the Clifford algebra $\{\gamma_\mu,\gamma_\nu\} = 2\delta_{\mu\nu}$ in 2+1 dimensions.
The matrix denoted $M$ is defined
\bea
\label{Mdef}
M = 
\left[\begin{array}{ccc}
~1~ & ~0~ & ~0~ \\
0 & v_F  & 0 \\
0 & 0 & v_F   \\
\end{array}
\right]\,.
\eea
Lorentz invariance is explicitly broken by the presence of this matrix with $v_F\ne 1$. 

The non-local nature of the gauge field action is due to the fact that the photon which mediates the interactions between the electrons propagates out of the graphene plane, in the bulk of the 3+1 dimensional space-time. 
This so called ``brane action'' can be obtained by integrating out the photon momentum modes in the third spatial dimension \cite{brane}. 

We include non-perturbative effects by introducing fermion dressing functions, and solving a set of coupled Dyson-Schwinger equations. 
We make a simple approximation for the photon polarization tensor, but within this approximation we keep all frequency dependence.
We use frequency dependent dressing functions and include vertex corrections (using an ansatz which is constructed to preserve gauge invariance).
We also include magnetic effects, and full frequency dependence in the loop integrals. 
In the language of the literature, this means dropping the Coulomb approximation and including retardation effects.

There are two important limitations to our approach. 
First, we use the one loop frequency dependent photon polarization tensor calculated with bare lines.
The vanishing of the density of states at the Dirac points makes this a reasonable approximation.
A complete calculation would include dressing functions for each independent component of the photon polarization. 
We use the one loop result for simplicity only - this is not a fundamental limitation of our method which could be extended in a straightforward way to include self-consistently determined photon dressing functions. However, such a calculation
would be extremely cost-intensive in terms of CPU-power and therefore we postpone relaxing this limitation to the future.

More importantly, the effective theory we are using is only valid at low energies, close to the Dirac points, where the fermions have linear dispersion relations. This means, for example, that we do not include the screening from higher energy quasi-particle states ($\sigma$-bands) by using an additional
momentum dependent factor $\epsilon(k)$ as in Refs.~\cite{wehling11,lorenz1}.

It is interesting to understand how our method is related to non-relativistic  approaches. 
In solid state physics, the electrons are usually assumed to interact through a long-range Coulomb interaction.
Because the photons that mediate the Coulomb interaction move so much more quickly than the electrons, it is usual to assume that the Coulomb interaction is instantaneous, and that magnetic interactions are suppressed. 
The Coulomb interaction is screened by vacuum  particle-hole pair production. This screening effect is normally included using a frequency dependent Lindhard screening function, which is a specific approximation to the electric part of the one loop vacuum photon polarization tensor discussed above. 
This calculation produces a strong renormalization of the fermi-velocity, which agrees with what is seen experimentally. The value of the critical coupling depends strongly on the precise form of the Lindhard screening function, which indicates an extreme sensitivity to the  approximation. 
We will show that the Coulomb approximation with frequency dependent Lindhard screening can be extracted from our more complete calculation as a specific well defined limit.

\section{Notation}
\label{notation-section}

We work in Landau gauge. The Euclidean space Feynman rules obtained from the action in (\ref{action}) are 
\bea
\label{bareFR}
&& S^{(0)}(P)=\big[i\gamma_\mu M_{\mu\nu} P_\nu\big]^{-1}\,,\\[2mm]
&& G^{(0)}_{\mu\nu}(Q)=\big[\delta_{\mu\nu}-\frac{Q_\mu Q_\nu}{Q^2}\big]\,\frac{1}{2\sqrt{Q^2}}\,, \\[1mm]
\label{barevert}
&& \Gamma^{(0)}_\mu = M_{\mu\nu}\gamma_\nu\,,
\eea
where we use the notation $Q_\mu = (q_0,\vec q)$ and $Q^2=q_0^2+q^2$, and similarly for the momenta $P$ and $K=P-Q$.

We include the non-perturbative effect of interactions by introducing dressing functions in the fermion and photon propagator. Since Lorentz invariance is explicitly broken, the fermion propagator contains three dressing functions which we call $Z(p_0,\vec p)$, $A(p_0,\vec p)$ and $\Delta(p_0,\vec p)$. Defining the diagonal 3$\times$3 matrix
\bea
\label{Amatrix}
{\bf A}(p_0,\vec p) = 
\left[\begin{array}{ccc}
Z(p_0,\vec p) & 0 & 0 \\
0 & A(p_0,\vec p)  & 0 \\
0 & 0 & A(p_0,\vec p)   
\end{array}
\right]\,,
\eea
the dressed fermion propagator has the form
\bea
\label{SF}
&& S^{-1}(P) = i \gamma_\mu {\bf A}_{\mu\nu}(p_0,\vec p) M_{\nu\tau}P_\tau + \Delta(p_0,\vec p)\,.
\eea
To obtain self-consistent expressions for the dressing functions we rewrite the inverse propagator as
\bea
\label{SF2}
&& S^{-1}(P) = (S^{(0)})^{-1}(P)+\Sigma(P)
\eea 
and use the Dyson equation to represent the fermion self energy
\bea
\label{fermion-SD}
&& \Sigma(p_0,\vec p) = e^2\int dK G_{\mu\nu}(q_0,\vec q)M_{\mu\tau}\gamma_\tau S(k_0,\vec k) \Gamma_\nu\,,
\eea
where we use the notation
\bea
\label{d4K}
dK = \int \frac{dk_0\,d^2k}{(2\pi)^3}\,,~~~~Q=P-K\,.
\eea
We obtain independent equations for each of the dressing functions by calculating the trace of (\ref{SF}) and (\ref{SF2}) multiplied by appropriate projection operators. These projectors are
\bea
\label{Zfactor}
&& \text{factor}_Z = \frac{i}{4 p_0}\gamma_0 \,,~~~
 \text{factor}_A = \frac{i}{4 v_F p^2}(P_\mu\gamma_\mu-p_0\gamma_0) \,,~~~
 \text{factor}_\Delta = \frac{1}{4} \,.
\eea

\begin{figure}[tb]
\begin{center}
\includegraphics[width=14cm]{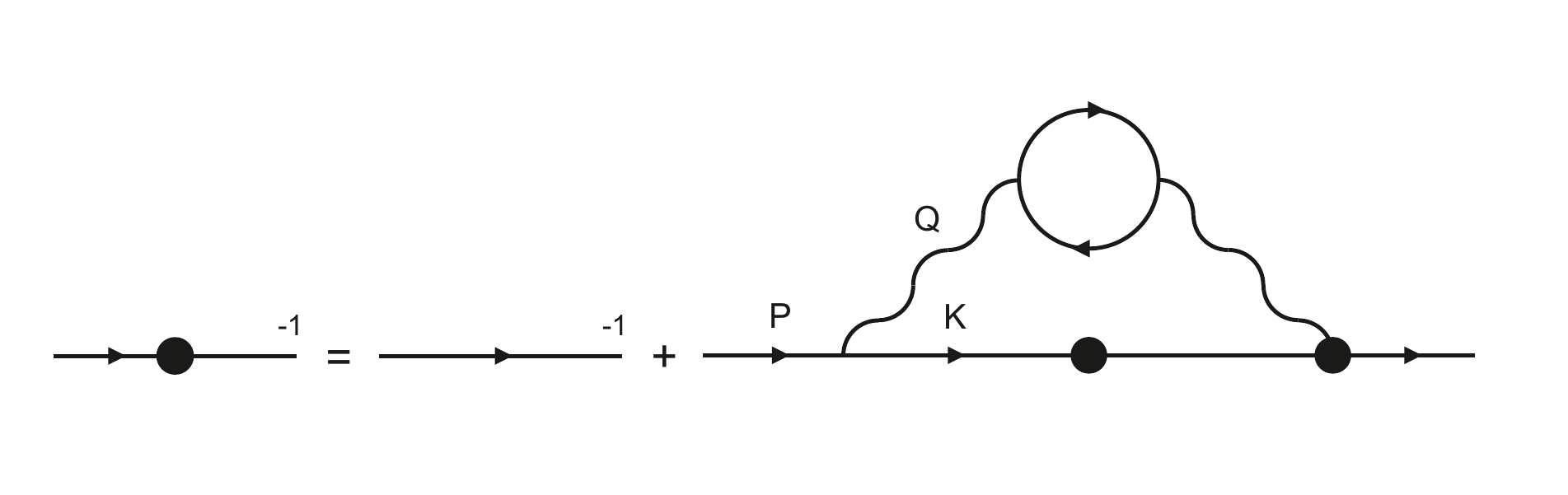}
\end{center}
\caption{The fermion Dyson-Schwinger equation in equation (\ref{SF2}) with self energy given in (\ref{fermion-SD}) and photon propagator in (\ref{fullG}-\ref{trpi}). The blobbed line and vertex represent, respectively, the dressed fermion propagator and the dressed vertex. \label{sdeqn-fig}}
\end{figure}

We use a 1-loop approximation for the photon polarization tensor. 
We define projection operators and decompose the polarization tensor (which we assume to be transverse),
\bea
&& P^1_{\mu\nu}=\delta_{\mu\nu}-\frac{Q_\mu Q_\nu}{Q^2}\,,~~P^2_{\mu\nu}=\frac{Q_\mu Q_\nu}{Q^2}\,,~~
P^3_{\mu\nu} = \frac{n_\mu n_\nu}{n^2}\,,~~n_\mu = \delta_{\mu 0}-\frac{q_0Q_\mu}{Q^2}\,,\\
&& \Pi_{\mu\nu}=\alpha P^1_{\mu\nu}+\gamma P^3_{\mu\nu}\,.
\eea
The inverse photon propagator in Lorentz gauge is written
\bea
G_{\mu\nu}^{-1} = \frac{2}{\sqrt{Q^2}}\big(Q^2 P_{\mu\nu}^1+\frac{1}{\xi} P^2_{\mu\nu}\big)+\Pi_{\mu\nu}\,.
\eea
Inverting this equation and choosing Landau gauge ($\xi=0$) the propagator is 
\bea
\label{fullG}
&& G_{\mu\nu} =  
\frac{P_{\mu\nu}^1}{G_T(q_0,\vec q)}+ P_{\mu\nu}^3\left(\frac{1}{G_L(q_0,\vec q)}-\frac{1}{G_T(q_0,\vec q)}\right)\,,\\[4mm]
&& G_T(q_0,\vec q) = 2\sqrt{Q^2}+\alpha\,,~~~G_L(q_0,\vec q) = 2\sqrt{Q^2}+\alpha+\gamma \,.
\eea
The components $\alpha$ and $\gamma$ of the polarization tensor are obtained from:
\bea
&& \Pi_{00} = \frac{q^2}{Q^2}(\alpha+\gamma)\,,~~~~{\rm Tr}\,\Pi = 2\alpha+\gamma\,,\\
\label{pi00}
&& \Pi_{00} = \frac{\pi  \alpha  q^2 v_F}{\sqrt{q^2 v_F^2+q_0^2}}\,,\\
\label{trpi}
&& {\rm Tr}\,\Pi = \frac{\pi  \alpha  v_F \left(2 \left(q^2 v_F^2+q_0^2\right) + q^2
   \left(1-v_F^2\right)\right)}{\sqrt{q^2 v_F^2+q_0^2}}\,.
\eea
There are different possible choices for the vertex function $\Gamma_\nu$ in (\ref{fermion-SD}), this is discussed in Section \ref{freq-section}. The self-consistent equation we are solving is shown schematically in Fig. \ref{sdeqn-fig}. 

In order to simplify the equations we define some short-hand notation that we will use from this point forward. 
\bea
\label{nota1}
&& Z_p = Z(p_0,\vec p)\,,~~A_p = A(p_0,\vec p)\,,~~\Delta_p = \Delta(p_0,\vec p)~~ \text{and likewise for }k\\
&& Z_s=Z_p+Z_k\,,~~~Z_d=Z_p-Z_k ~~~ \text{and likewise for $A$ and $\Delta$} \\[2mm]
&& G_L=G_L(q_0,\vec q)\,,~~~ G_T=G_T(q_0,\vec q)\,,~~~S_k = k_0^2 Z_k^2 + k^2 A_k^2 v_F^2+\Delta _k^2\,,\\[2mm]
&& dK = \int \frac{dk_0\,d^2k}{(2\pi)^3}\,,~~~x=\frac{\vec p\cdot\vec k}{p\,k}\,.
\eea
We set $\hbar=c=1$.

\subsection{Frequency independent dressing functions} 
\label{nofreq-section}
We start with a simple calculation that involves several assumptions:
\begin{enumerate}
\item We assume that we can use bare vertices (we discuss the consequences for gauge invariance below). 
\item If the contributions from the transverse (magnetic) modes of the photon propagator are suppressed, then we need only to include the 00 component of the photon propagator so that $G_{\mu\nu} = \delta_{\mu 0}\delta_{\nu 0}G_{00}$ with
\bea
&& \label{G00}
G_{00}(q_0,q) = \frac{q^2}{Q^2(\alpha+\gamma+2\sqrt{Q^2})} \equiv \frac{1}{f}\;\frac{1}{2\sqrt{f q^2}+f\Pi_{00}}\,,~~~f = \frac{Q^2}{q^2}\,.
\eea
From the structure of the bare vertex (see equations (\ref{Mdef}) and (\ref{barevert})) it appears that this assumption is equivalent to using $v_F\ll 1$. This is not quite true however, because at leading order in $v_F$ there are contributions to $A_p$  when $\Gamma_0^{(0)}= 1$ couples to $v_F A_k$, and when $\Gamma^{(0)}_i =  v_F$ couples to $Z_k$.
\item The dressing functions are taken to be independent of frequency (we do not introduce different notation but simply use the same symbols as in equation (\ref{nota1}) which are now taken to mean $Z_p=Z(p)$, etc).
\end{enumerate}

Using these three assumptions the integrals for $Z$, $A$ and $\Delta$ have the simple form 
\bea
&& Z_p = 1 - \frac{4\alpha\pi v_F}{p_0} \int dK \,  \,k_0 \; G_{00}(q_0,\vec q)\;\frac{Z_k}{S_k}\,,\nonumber\\
\label{nofreq-1}
&& A_p = 1 + \frac{4\alpha\pi v_F}{p^2} \int dK \,  \, \vec k \cdot \vec p \; G_{00}(q_0,\vec q)\;\frac{A_k}{S_k}\,,\\
&&D_p =  4\alpha\pi v_F \int dK \, \,G_{00}(q_0,\vec q)\; \frac{\Delta _k}{S_k}\,.\nonumber
\eea
We refer to this calculation as \underline{$\omega$-independent-full} (`$\omega$ independent' because the dressing functions are assumed to be independent of frequency, and the word `full' indicates the absence of assumption (4), which is given below).\\

It is common to make the following additional assumption:
\begin{enumerate}
\setcounter{enumi}{3}
\item 
The photons that mediate the Coulomb interaction move much more quickly than the the electrons, and therefore the interaction can be taken to be almost instantaneous, which means $q_0\ll q$.
\end{enumerate}
Since both $v_F$ and $q_0$ are small we should consider the relative size of the two small parameters. 
We use the approximation that both parameters are the same order: $\{v_F,q_0\}\sim \delta$ and work to leading order in $\delta$. 
This means that we set $f=1$ in (\ref{G00}) but do not drop the $q_0$ dependence in the polarization tensor  (\ref{pi00}), which gives the static (or Coulomb) propagator
\bea
\label{staticG}
 G_{00}^{\rm sttc}(q_0,\vec q) := G_{00}(q_0,\vec q)\big|_{f=1}\,.
\eea
The first equation in (\ref{nofreq-1}) gives $Z=1$ and the equations for $A$ and $\Delta$ are obtained by replacing $G_{00}$ with $G_{00}^{\rm sttc}$. 
Note that $Z=1$ means that (1) and (2) together satisfy the Ward identity, but if (1) is used without (2) the Ward identity is not satisfied.
Using this leading order in $\delta$ approximation, the frequency integrals can be done analytically.
We refer to this calculation as \underline{$\omega$-independent}. 
It was done in Ref. \cite{pop} and produces a critical $\alpha$ that is $\sim 7.8$.

Notice that assumptions (3) and (4) together imply a restriction on the frequencies of the virtual particles in the loop.
If the structure functions are independent of frequency (assumption (3)) then we can choose $p_0=0$ in all integrals. 
This gives $q_0=-k_0$ and therefore if $q_0$ is small (assumption (4)) then $k_0$ is necessarily also small. 
These two assumptions are therefore not compatible, unless the frequency integral is dominated by the small $k_0$ region. 
However, the authors of Ref. \cite{pop} did the same calculation using the instantaneous approximation (which means $\Pi_{00}(k_0,\vec q) \to \Pi_{00}(0,\vec q) = \alpha\pi q)$ and found that no gap is obtained for arbitrarily large values of the effective coupling ($\alpha_c\to\infty$). This result indicates that the way in which the frequency dependence of the integrand is approximated can have a crucial effect on the result. 

A comparison of the $\omega$-independent-full and $\omega$-independent calculations provides a useful check of the procedure. If assumption (4) is justified, the two calculations should give approximately the same critical alpha. Some details of the calculation are given in Section \ref{results-section}.
We obtain $\alpha_c = 7.80$ from the $\omega$-independent calculation, and the $\omega$-independent-full calculation gives $\alpha_c=8.95$. The increase in $\alpha_c$ which results when we do the full frequency integral instead of using the Coulomb propagator indicates that the net effect of the $\delta$ expansion is to decrease screening. 
These values are compared with results from our calculations with frequency dependent dressing functions in Section \ref{results-section}.

\subsection{Frequency dependent dressing functions}
\label{freq-section}
The assumptions that are made in Section \ref{nofreq-section} are interconnected, and therefore we cannot relax them one at a time. In this section we drop all of the assumptions discussed previously: we use frequency dependent dressing functions,  include all components of the photon propagator, and use a non-perturbative self-consistently determined vertex. The only approximation we make is to drop terms of order $v_F^3$ relative to terms of order $v_F$ (we have checked that the contribution from the $v_F^3$ terms is negligible). 

The three frequency dependent dressing functions are written $Z(p_0,\vec p)$, $A(p_0,\vec p)$ and $\Delta(p_0,\vec p)$, and sometimes abbreviated as in equation (\ref{nota1}).
We assume that they are even functions of the frequency variable. 
In order to preserve gauge invariance, we define a non-covariant extension of the Ball-Chiu vertex \cite{ball-chu}:
\bea
\label{BC}
\Gamma_\mu && = \frac{1}{2}\big({\bf A}_{\mu\nu}(p_0,\vec p)+{\bf A}_{\mu\nu}(k_0,\vec k)\big)\gamma_\nu\\
&&+\left[\frac{1}{2}(P_\sigma+K_\sigma)\big({\bf A}_{\sigma\nu}(p_0,\vec p)-{\bf A}_{\sigma\nu}(k_0,\vec k)\big)\gamma_\nu 
+ i \big(\Delta(p_0,\vec p)-\Delta(k_0,\vec k)\big)\right]\frac{(P_\mu+K_\mu)}{P^2-K^2}\nonumber
\eea
which satisfies the Ward identity
\bea
\label{wi}
-i Q_\mu\Gamma_\mu = S^{-1}(p_0,\vec p) - S^{-1}(k_0,\vec k)\,.
\eea
In addition to difficulties associated with the vastly increased phase space, the calculation is tricky because of the terms in the vertex of the form 
\bea
\label{nastyX}
\frac{X(p_0,\vec p)-X(k_0,\vec k)}{P^2-K^2}\,,~~~X\in\{Z,A,\Delta\}\,,
\eea
which approach 0/0 $\to$ constant as $K\to P$. We deal with this problem by constructing arrays of the independent variables that approach $P$ from both sides, but never touch the point $K=P$. 
We call this the \underline{BALL-CHIU} calculation, and we compare the results with two simpler versions which are described below.

Much simpler expressions are obtained if we use the first term in the Ball-Chiu vertex (\ref{BC}) and expand in the parameter $\delta$ (as explained in the previous section). 
The transverse modes drop out and the longitudinal part of the propagator is replaced with the Coulomb propagator given in equation (\ref{staticG}).
In addition, the simpler ansatz for the vertex means that terms of the form shown in equation (\ref{nastyX}) do not appear. We refer to this as the \underline{COULOMB} 
calculation. It was done previously in Ref. \cite{rong-liu}. 

Another possibility is to use the first term in the Ball-Chiu vertex (\ref{BC}) but not the $\delta$ expansion, which means using the full longitudinal photon propagator including all momentum dependence instead of the static Coulomb approximation.
We call this the \underline{SHORT} calculation (which refers to the fact that the only approximation is the 
shortened vertex ansatz).


Using the first term in the Ball-Chiu vertex we obtain the relatively simple expressions
\bea
&& Z_p = 1-\frac{2\alpha\pi v_F}{p_0 } \int dK \frac{k_0 q^2 Z_k Z_s}{Q^2\,G_L S_k} \,,\\
\label{shortA}
&& A_p = 1 + \frac{2\alpha\pi v_F}{p^2} \int dK \; \frac{q^2 A_k Z_s \vec k \cdot \vec p  +  k_0 q_0 Z_k(Z_s+A_s) \vec p \cdot \vec q }{Q^2\,G_L S_k}\,,\\
&&D_p =  2\alpha\pi v_F \int dK \frac{q^2 \Delta _k Z_s}{Q^2\,G_L S_k}\,.
\eea
These integrals correspond to the SHORT calculation. To obtain the COULOMB expressions we drop the second term in the numerator of (\ref{shortA}), set the factor $q^2/Q^2$ in each remaining integrand to one, and replace $1/G_L(q_0,q)$ with the static Coulomb propagator defined in (\ref{staticG}). 
Using the full Ball-Chiu vertex we obtain the integrands for the BALL-CHIU calculation
\bea
&& Z_p = 1 -  \frac{2\alpha\pi v_F}{p_0}  \int dK \frac{k_0}{Q^2 \,G_L S_k} \\
&&~~~~~ \left[q^2 Z_k Z_s  -\frac{Z_k \left(q_0 \left(k^2-p^2\right) \left(k_0+p_0\right)+q^2
   \left(k_0+p_0\right){}^2\right) Z_d}{K^2-P^2}  \right]\nonumber\\
&& A_p = 1 +  \frac{2\alpha\pi v_F}{p}  \int dK  \frac{1}{Q^2\,S_k}\\
&&~~~~~ \left[
\frac{k q^2 x A_k Z_s}{G_L}
+
\frac{k_0 q_0 Z_k \left(A_s+Z_s\right) (p-k x)}{G_L} 
-\frac{2 k^2 k_0 p Q^2 \left(x^2-1\right) Z_d \left(k_0+p_0\right)
   Z_k}{G_T q^2 \left(K^2-P^2\right)} \right. \nonumber\\
&&~~~~~\left. + \frac{\left(q_0 \left(k^2-p^2\right)+k_0 q^2+p_0 q^2\right) \left(k_0 q^2 A_d
   Z_k (k x+p)-Z_d \left(k_0+p_0\right) \left(k q^2 x A_k+k_0 q_0 Z_k (p-k
   x)\right)\right)}{G_L q^2 \left(K^2-P^2\right)}
\right]\nonumber\\
&& D_p = 2\alpha\pi v_F  \int dK \frac{1}{Q^2 \,G_L S_k} \\
&&~~~~~ \left[q^2 \Delta _k Z_s
+ \frac{2 k_0 \Delta _d Z_k \left(q_0 \left(k^2-p^2\right)+k_0 q^2+p_0
   q^2\right)}{K^2-P^2}
-\frac{Z_d \Delta _k \left(k_0+p_0\right) \left(q_0 \left(k^2-p^2\right)+k_0
   q^2+p_0 q^2\right)}{K^2-P^2}  \right]\,.\nonumber
\eea

\section{Results}
\label{results-section}

To do the $k$-momentum integral, we introduce an ultra-violet cut-off $\Lambda$ and use a logarithmic scale, in order to increase the sensitivity of the numerical integration procedure to the infra-red regime where the dressing functions change most rapidly.
The integral over the frequency $k_0$ can also be done using a logarithmic scale, but we would like to consider the possibility that the upper limit of the frequency integral ($\Lambda_0$) should not necessarily be taken to be the same as that of the momentum integral ($\Lambda$). 
Physically we expect that the electrons will not respond to photons at very high frequencies, and this effect should be taken into account by the non-covariant form of the integrand, which tells us that frequencies $k_0\sim v_F=1/300$ should contribute with the same importance as momenta $k\sim 1$.
Naively therefore, it seems that we should take the upper limit of both integrals as equal. 
We have checked this by using a compactified frequency variable $k_0=y^3/(1-y^2)$ and integrating over $y$ from zero to one. Results from the two procedures are given in Table \ref{table-alphac}.

We define dimensionless variables $\hat k_0=k_0/\Lambda$, $\hat p_0=p_0/\Lambda$, $\hat k=k/\Lambda$, $\hat p=p/\Lambda$ and $\hat\Delta =\Delta/\Lambda$. The hatted frequency and momentum variables range from zero to one. 
From this point on, we suppress all hats.  

For convenience we give below a list of the calculations we have done, and the physical content of each one. 
\begin{enumerate}
\item 
$\omega$-independent:  frequency independent dressing functions, no retardation effects
\item
$\omega$-independent-full:   frequency independent dressing functions, retardation effects included  
\item
COULOMB:  frequency dependent dressing functions, first term in the Ball-Chiu vertex, Coulomb approximation, no retardation effects
\item
SHORT: frequency dependent dressing functions, first term in the Ball-Chiu vertex, no magnetic contributions, retardation effects included
\item
BALL-CHIU:    frequency dependent dressing functions, full Ball-Chiu vertex, electric and magnetic contributions, retardation effects included
\end{enumerate}

\begin{figure}[h]
\centering
\mbox{\subfigure[]{\includegraphics[width=3.2in]{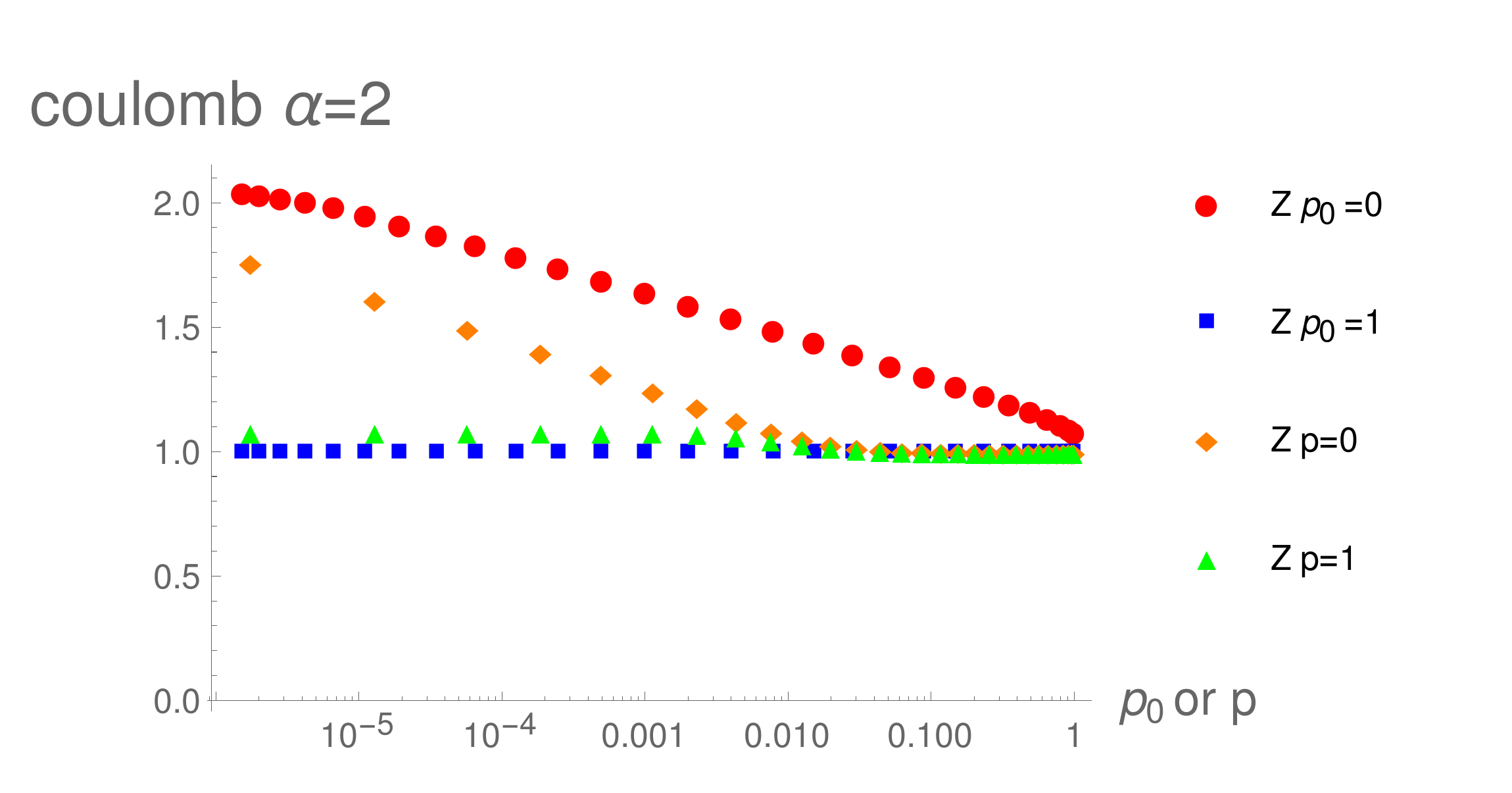}}\quad
\subfigure[]{\includegraphics[width=3.2in]{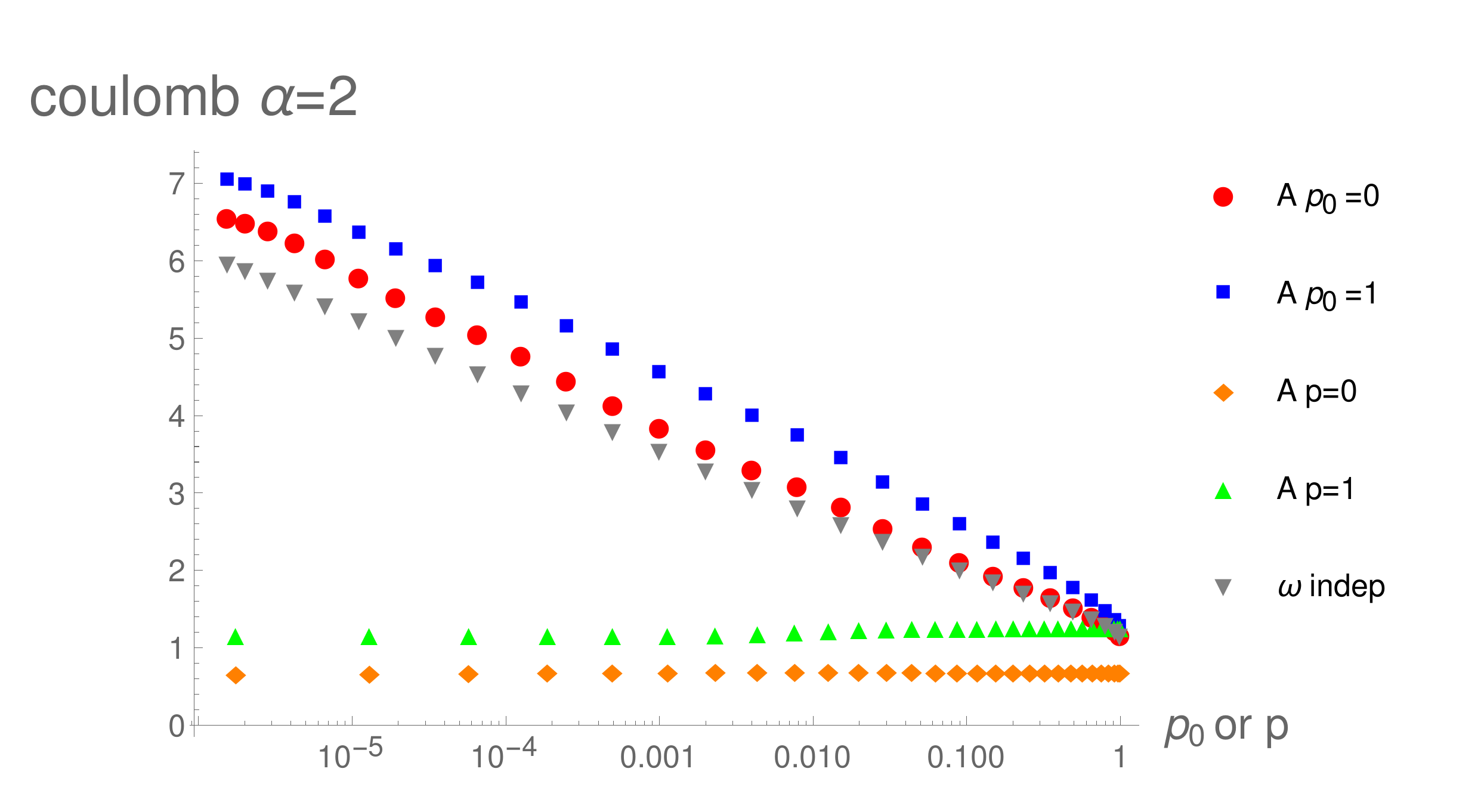} }}
\mbox{\subfigure[]{\includegraphics[width=3.2in]{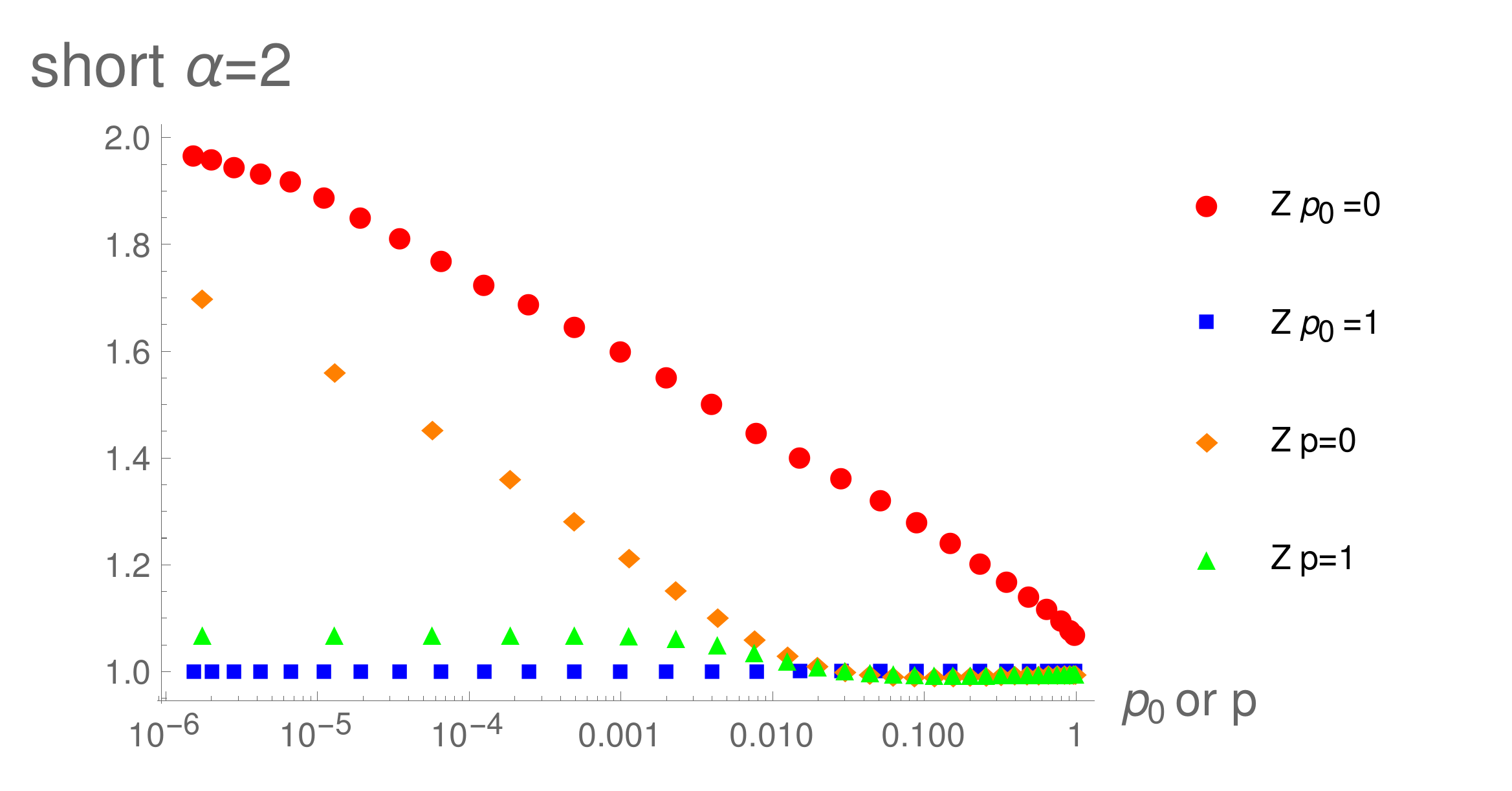}}\quad
\subfigure[]{\includegraphics[width=3.2in]{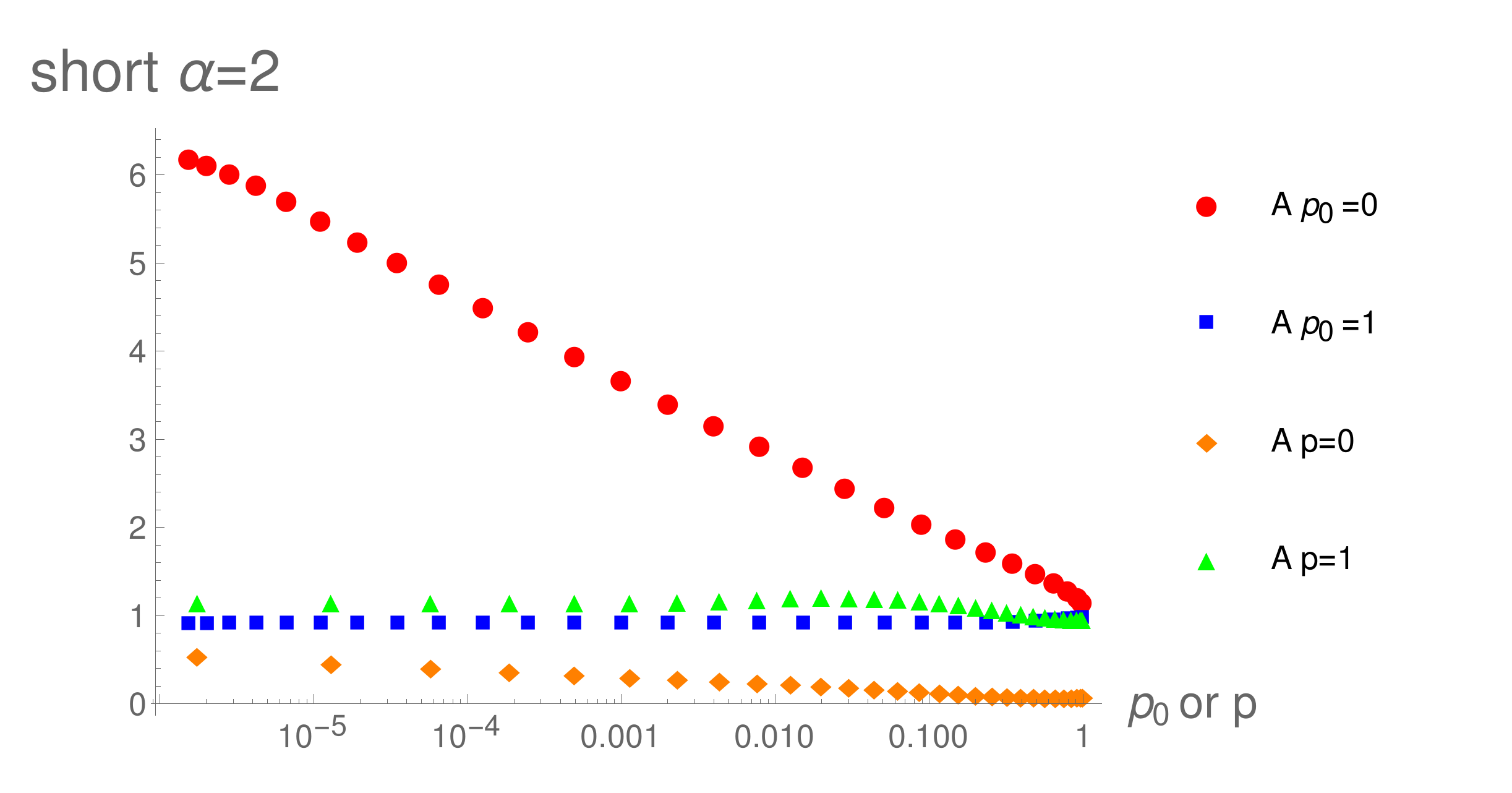} }}
\mbox{\subfigure[]{\includegraphics[width=3.2in]{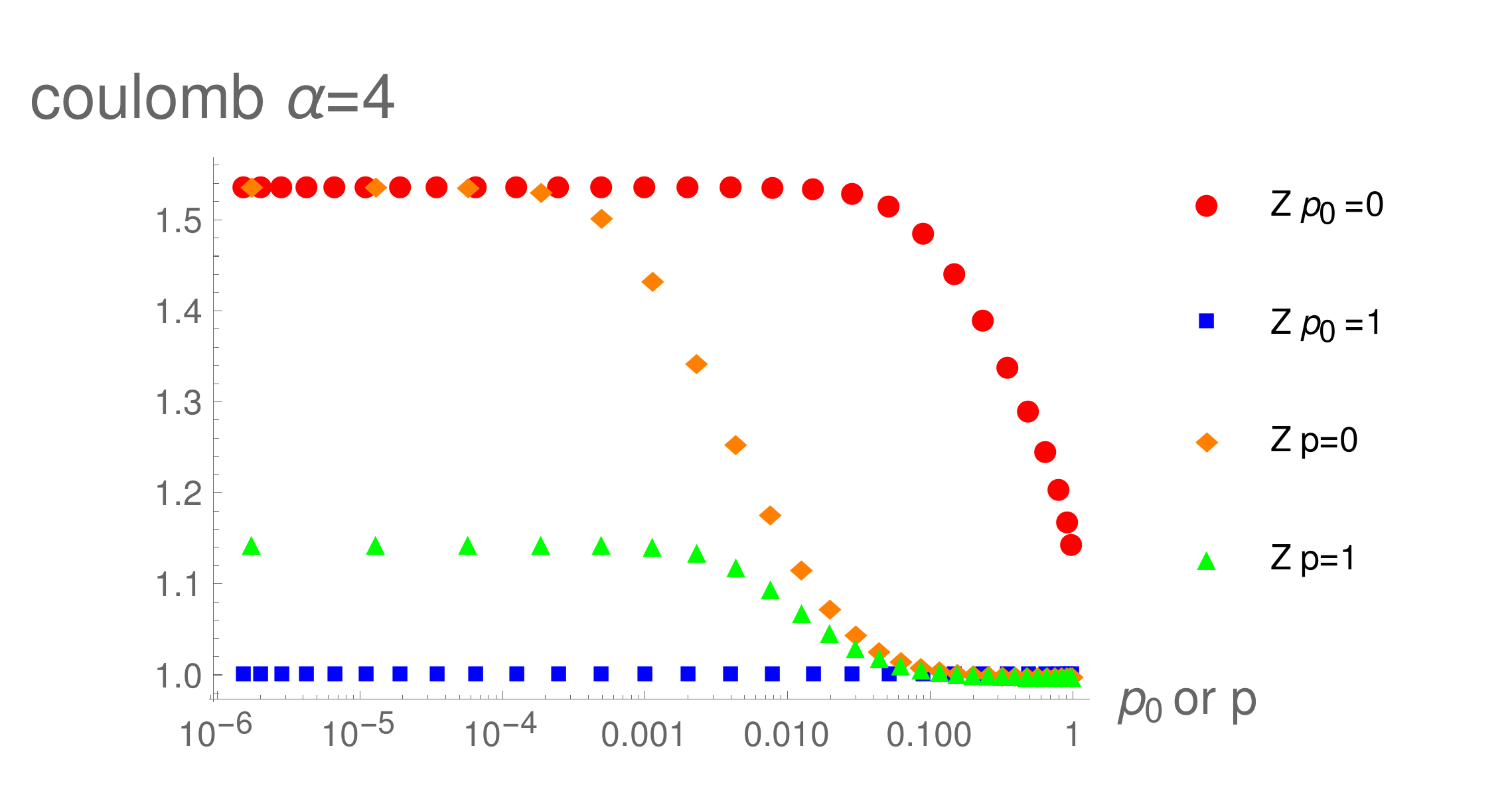}}\quad
\subfigure[]{\includegraphics[width=3.2in]{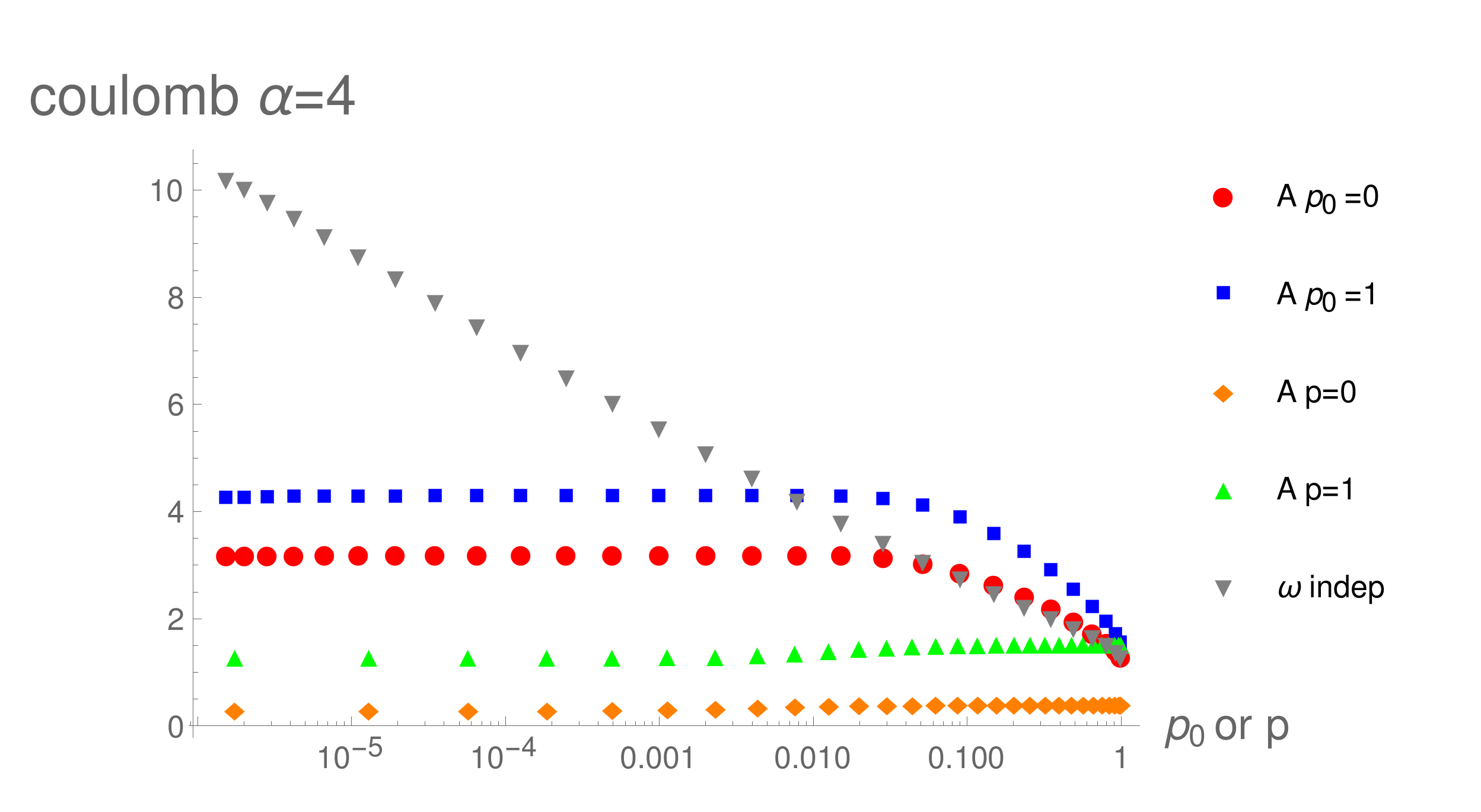} }} 
\mbox{\subfigure[]{\includegraphics[width=3.2in]{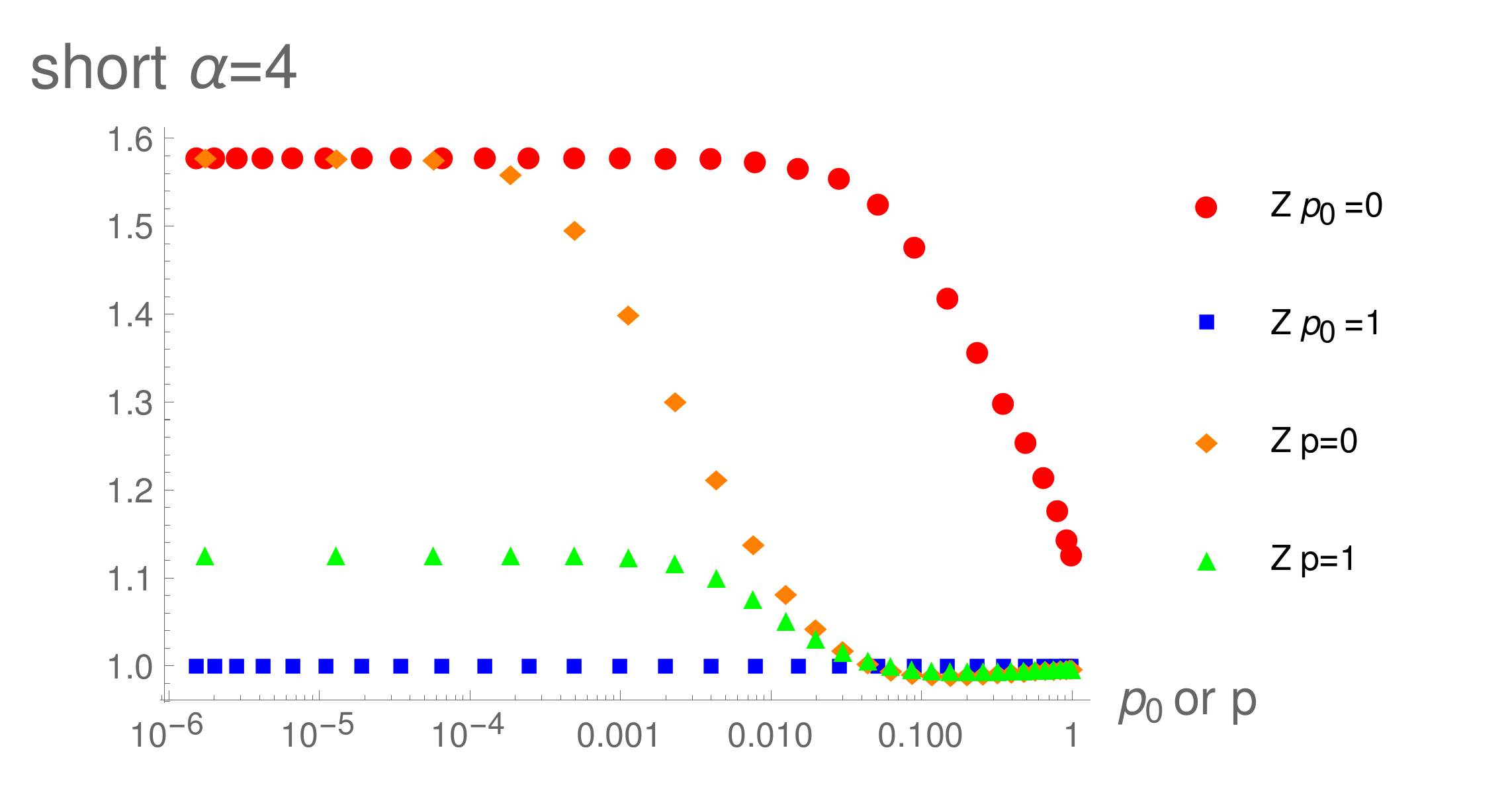}}\quad
\subfigure[]{\includegraphics[width=3.2in]{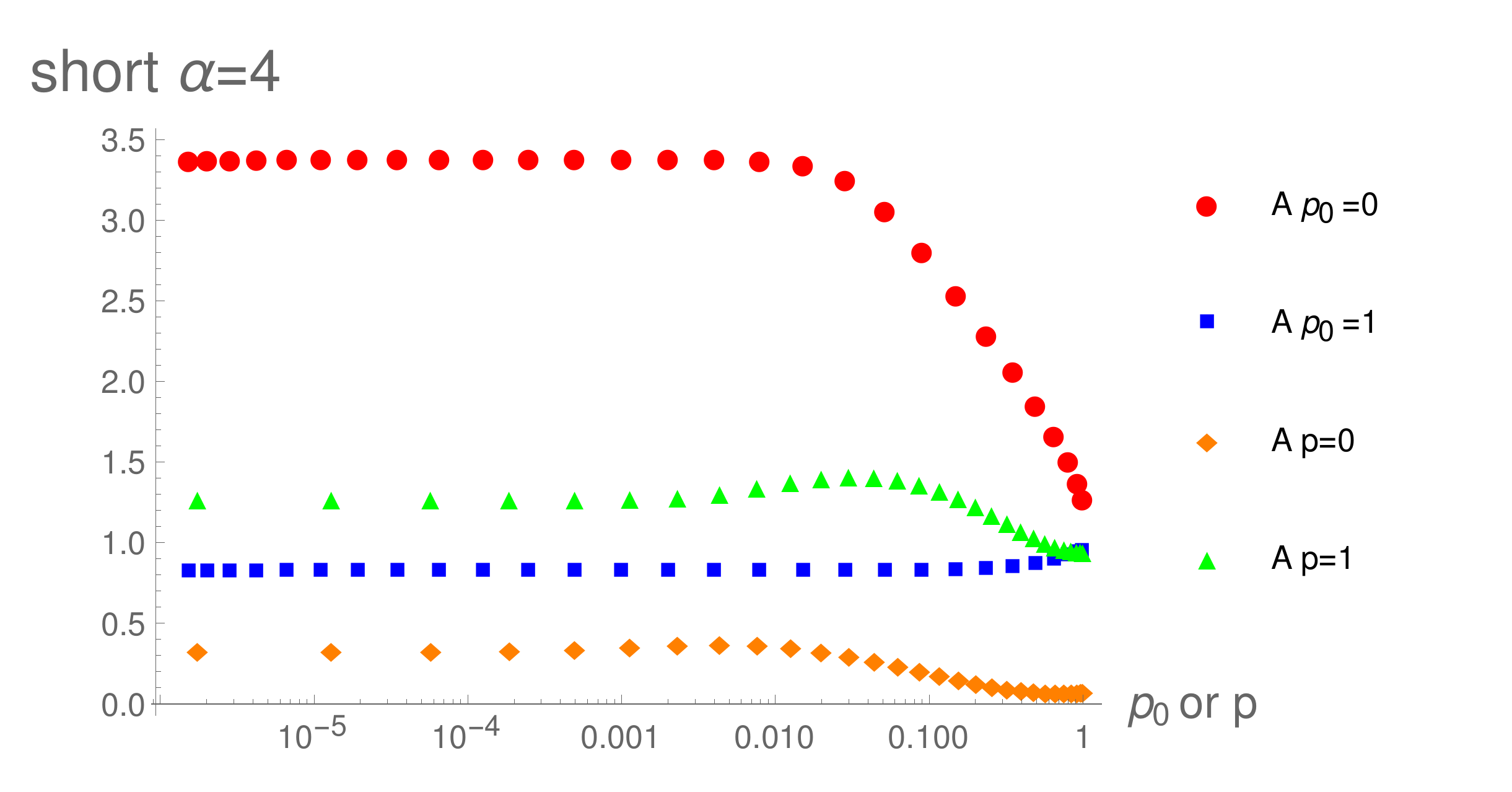} }}
\caption{Momentum dependence of the $Z$ and $A$ dressing functions. \label{ZA-mom-fig}}
\end{figure}

\begin{figure}[t]
\centering
\mbox{\subfigure[]{\includegraphics[width=3.2in]{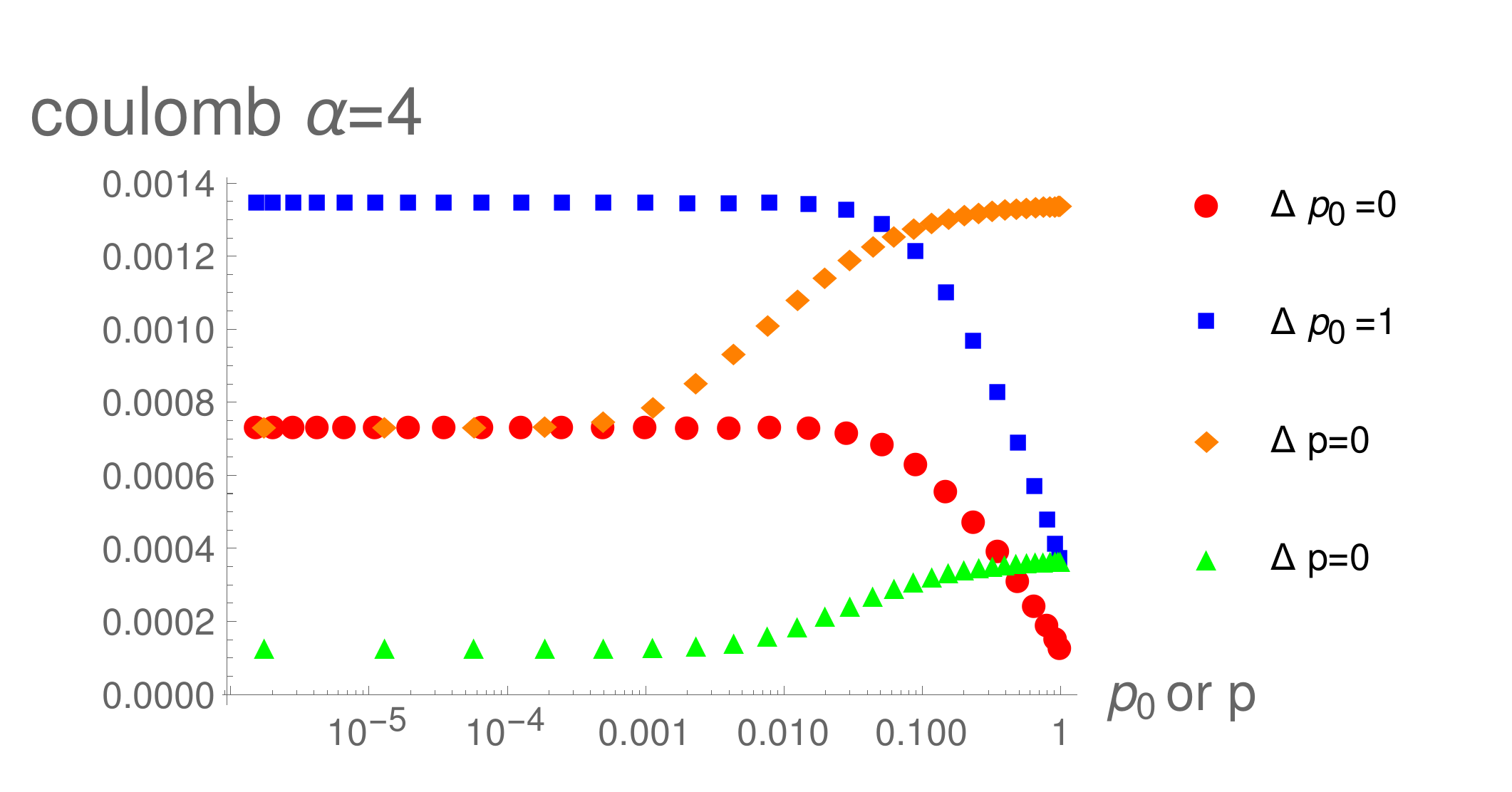}}\quad
\subfigure[]{\includegraphics[width=3.2in]{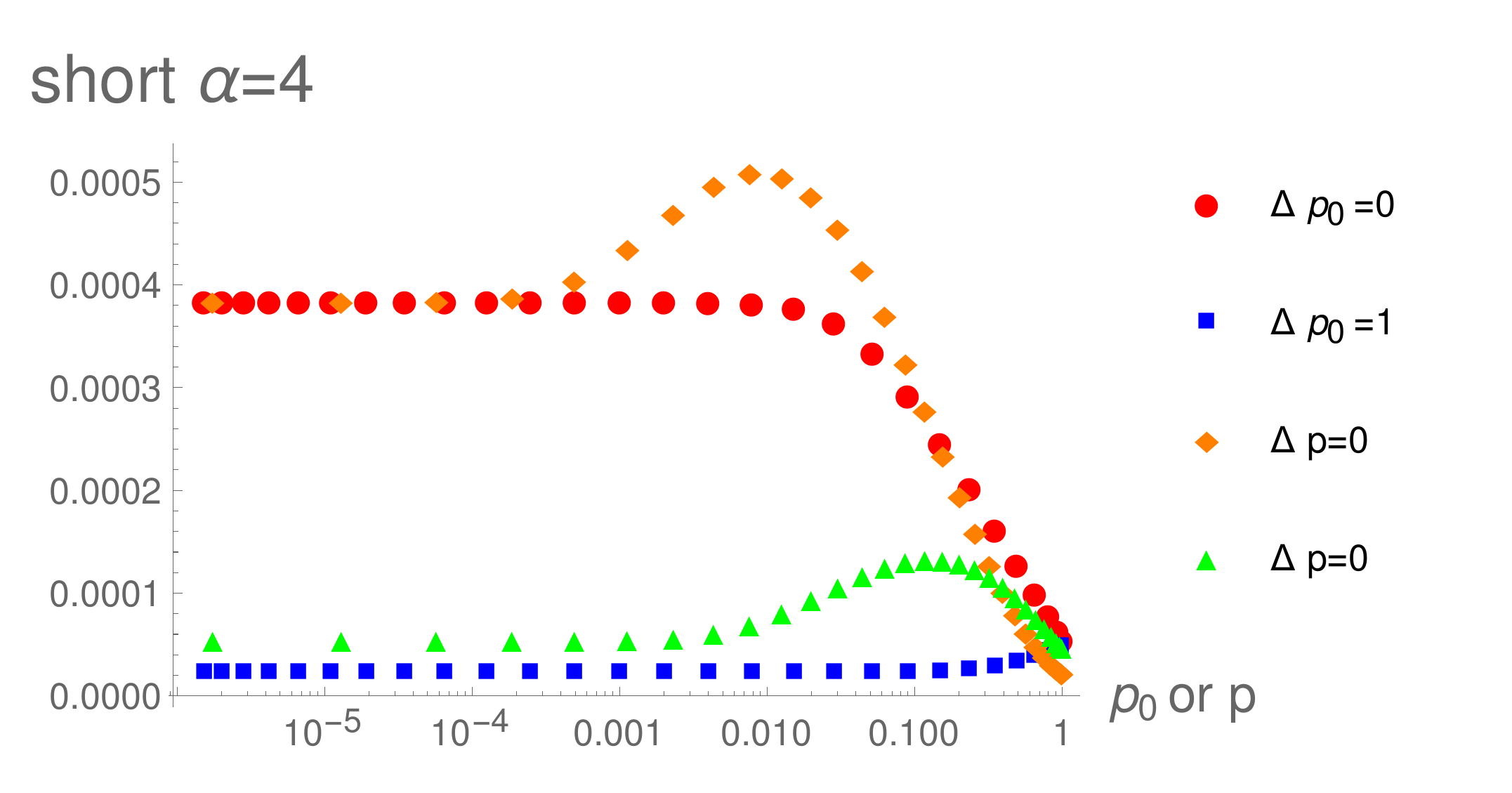} }}
\caption{Momentum dependence of the $\Delta$ dressing function. \label{D-mom-fig}}
\end{figure}
\begin{figure}[b]
\begin{center}
\includegraphics[width=10cm]{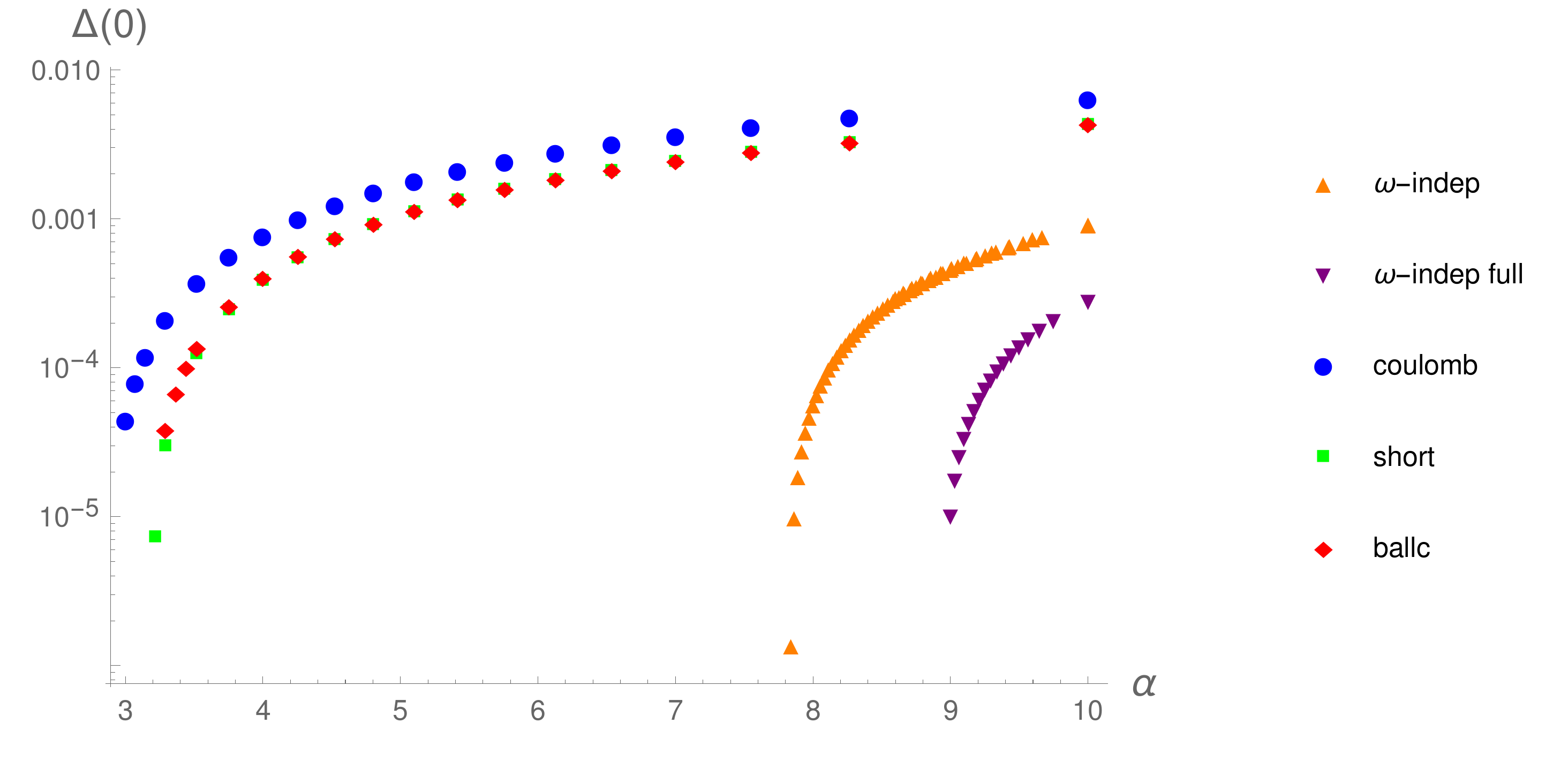}
\end{center}
\caption{$\hat\Delta$ versus $\alpha$. The extrapolated values of the critical coupling are given in Table \ref{table-alphac}.  \label{alphaC-fig}}
\end{figure}
\begin{figure}[b]
\centering
\mbox{\subfigure{\includegraphics[width=3.2in]{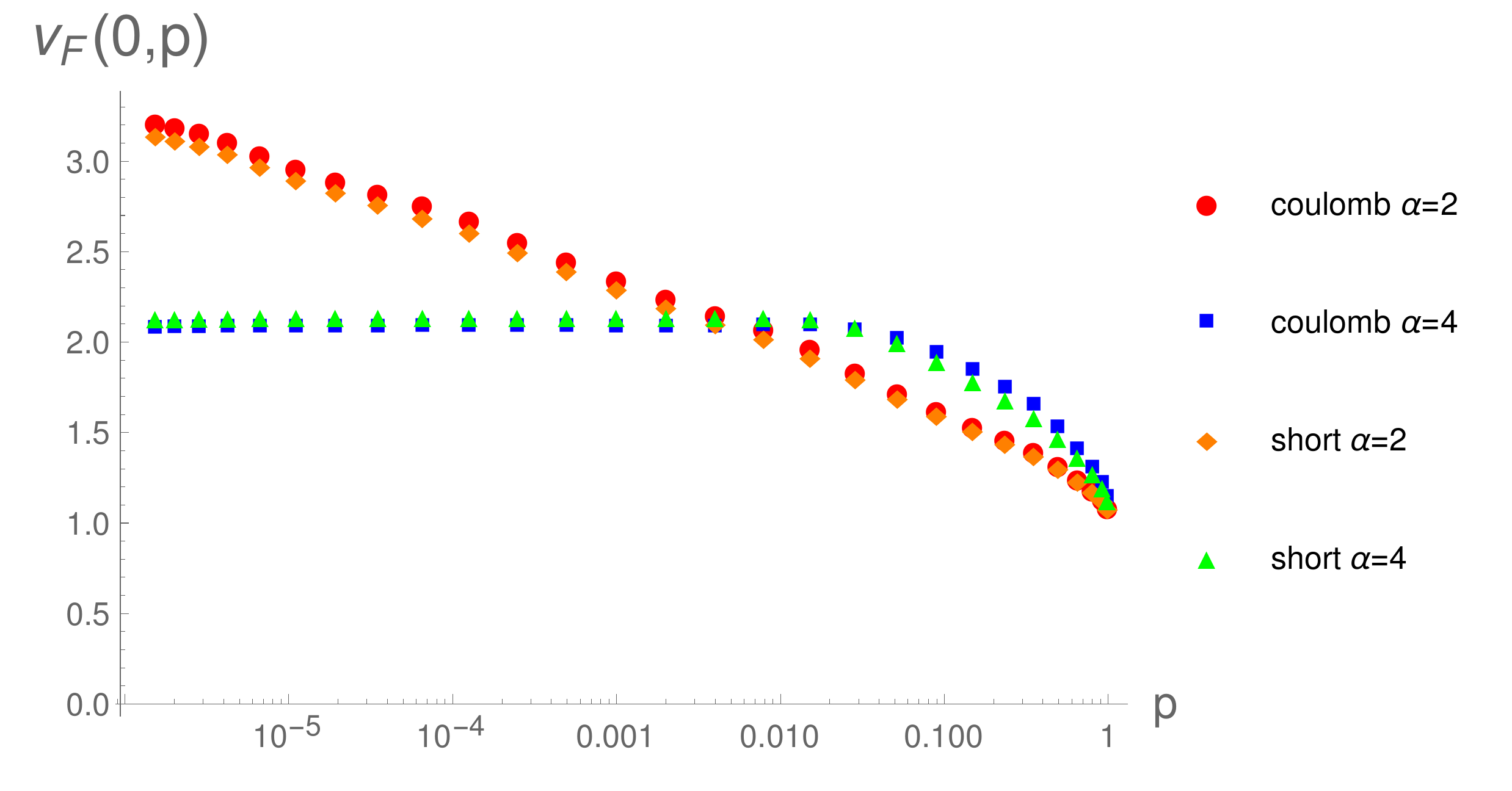}}\quad
\subfigure{\includegraphics[width=3.2in]{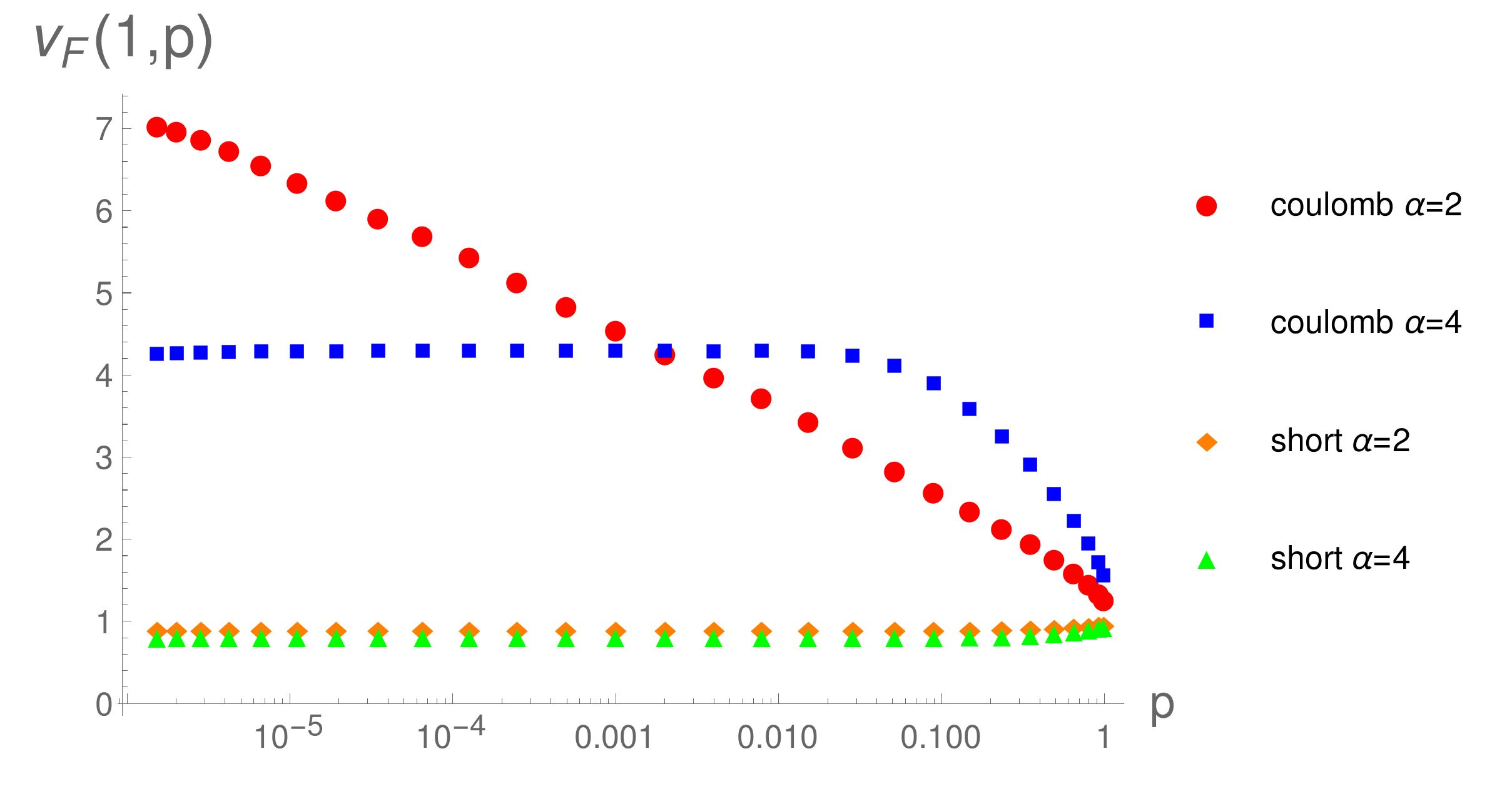} }}
\caption{The renormalized fermi velocity. \label{vF-fig}}
\end{figure}

In Figs. \ref{ZA-mom-fig} and \ref{D-mom-fig} we show the  momentum dependence of the dressing functions for different calculations and different values of $\alpha$. 
The results from the BALL-CHIU calculation are very close to those of the SHORT calculation, and therefore we do not show any BALL-CHIU plots. 
For both the COULOMB and SHORT calculations, the results obtained with $\Lambda_0=\Lambda$ and $\Lambda_0\to\infty$ are numerically almost identical. We show only results obtained using $\Lambda_0=\Lambda$.

We show each dressing function versus either $p$ or $p_0$. The variable which is not plotted is fixed at either the smallest or largest value available, which are $10^{-6}$ and 1. 
We use two different values of $\alpha$: $\alpha=2$, which is slightly below the critical value for all three of the calculations discussed in Section \ref{freq-section}, and $\alpha=4$, which is slightly above.
For $\alpha=2$ there are no plots of $\Delta$ since it is zero below the critical coupling. 
Each curve in the 10 graphs in Figs. \ref{ZA-mom-fig} and \ref{D-mom-fig} corresponds to one choice of 
calculation (COULOMB or SHORT); $\alpha$ (2 or 4); dressing function ($Z$, $A$ or $\Delta$); and value of the variable which was not plotted but held fixed ($p_0 =10^{-6}$, $p_0 =1$, $p =10^{-6}$ or $p =1$).

Comparing parts (a,c), (b,d), (e,g) and (f,h) we see that the COULOMB and SHORT calculations of $Z$ and $A$ for the same value of $\alpha$ agree fairly well in the infrared (the red and orange curves). We note however that although the shape of the curves is very similar, there are small but potentially important differences in the scale. 
In addition, there are significant differences in the ultraviolet (the green and blue curves). 
The critical coupling is the value of $\alpha$ for which the gap $\Delta(0,0)$ goes to zero in the infrared, which one expects to depend primarily on the infrared behavior of all three dressing functions. However, the self-consistent nature of the calculation makes it hard to estimate the effect of the ultraviolet regime. 

In Fig. \ref{ZA-mom-fig}(b) and \ref{ZA-mom-fig}(f) we show the momentum dependence of the dressing function $Z$ from the $\omega$-independent calculation, which can be compared with the result from the COULOMB calculation. In Fig. \ref{ZA-mom-fig}(b) the curve from the $\omega$-independent calculation is very close to both the $p_0$ large and $p_0$ small COULOMB curves, which indicates that the assumption of frequency independent dressing functions is quite reasonable at this value of $\alpha$. However, in Fig. \ref{ZA-mom-fig}(f) we see that the $\omega$-independent calculation is very far from the COULOMB one, and we therefore do not expect that the two calculations will produce the same critical coupling. 

In Fig. \ref{D-mom-fig} we see that the value of the gap at $\alpha=4$ is about twice as big in the COULOMB calculation, relative to the SHORT one. 
We expect therefore that the gap will disappear at a smaller coupling in the COULOMB calculation, which corresponds to a smaller critical coupling. 

In Fig. \ref{alphaC-fig} we show the value of the gap versus $\alpha$ for the five different calculations we have done. 
Extrapolation gives the values of the critical coupling shown in the third column of Table \ref{table-alphac}.
The numerical fits are done using {\it Mathematica}. We use three different methods, Spline, Hermite, and Automatic, and the results are the same to three decimal places in each case. 
As expected from the momentum data, $\alpha_c$ for the COULOMB calculation is slightly smaller than the SHORT result. 
Similarly, the $\omega$-independent calculation gives a slightly smaller $\alpha_c$ than the $\omega$-independent-full one. 
However, both of the frequency independent calculations give significantly larger results than the calculations which take into account the frequency dependence of the dressing functions. 

In Fig. \ref{vF-fig} we show the renormalized fermi velocity, which is defined as $A(p_0,p)/Z(p_0,p)$, as a function of momentum, at large and small frequency. The increase in the fermi velocity at small coupling which is observed experimentally is clearly seen. \\


We can also calculate the critical value of the coupling using a bifurcation analysis. 
We set $\Delta(p_0,p)=0$ and solve the coupled set of equations for $Z(p_0,p)$ and $A(p_0,p)$. 
Using these solutions, we attempt to solve self-consistently the equation for $\Delta(p_0,p)$ starting from an initial value of $10^{-12}$. 
For values of $\alpha$ above the critical coupling, the solution moves away from zero. Reducing $\alpha$, we search for the largest value for which the zero solution is stable. 
We have checked that this solution is independent of the initialization for $\Delta(p_0,p)$.
Using this method, we do not obtain any information about the momentum dependence of the dressing functions.
The advantage is that the calculation is much faster, for two reasons. Firstly, the total number of iterations is much smaller. If the number of iterations required to converge one dressing function is of order $N$, then the full calculation (which solves self-consistently 3 coupled equations for the functions $Z$, $A$ and $\Delta$) requires $\sim N^3$ iterations. The bifurcated calculation solves the equations in two steps and requires only $\sim N^2+N$ iterations. 
In addition, in the full calculation, the number of iterations increases as the critical point is approached. In the bifurcated calculation this critical slowing down does not appear. The bifurcation method therefore allows us to efficiently test the stability of the result for $\alpha_c$ against increases in the number of grid points used in the numerical integrals. The results for the upper and lower bounds of the critical coupling are shown in the fourth column 
of Table \ref{table-alphac}.

Our results from the $\omega$-independent calculation agree with \cite{pop} within the accuracy which is attainable 
from the fitting procedure. Our results from the COULOMB calculation differ slightly with \cite{rong-liu} where a value of $3.2<\alpha_c<3.3$ was reported.

\begin{table}[t]
\begin{center}
\begin{tabular}{|c|c|c |c|} 
\hline
 ~~~~~~~~ calculation ~~~~~~~~ & ~~~~~~~~ $k_0^{\rm max}/\Lambda$ ~~~~~~~~ & ~~~~~ $\alpha_c$ ~~~~~ & ~~~~~ bifurcation range ~~~~~ \\\hline\hline
$\omega$-independent & $\infty$ & 7.80 & 7.776-7.777\\
$\omega$-independent & $1$ & 8.967 & \\
$\omega$-independent-full  & $\infty$ & 8.955 &  \\ \hline
COULOMB  & $\infty$ & 2.889  & 2.882 - 2.880 \\ 
COULOMB  & $1$ & 2.906  & 2.900 - 2.899 \\ \hline 
SHORT  & $\infty$ & 3.189 & 3.188 - 3.190 \\ 
SHORT  & $1$ & 3.190  & 3.190 - 3.191 \\ \hline
BALL-CHIU  & $\infty$ & 3.178  & \\ 
BALL-CHIU  & $1$ & 3.178 &  \\ \hline
\end{tabular}
\end{center}
\caption{Results for critical values of the coupling $\alpha$. \label{table-alphac}}
\end{table}

\section{Conclusions}

We have done a calculation of the dynamically generated gap in mono-layer suspended graphene, starting from a low energy effective field theory.
We use a non-perturbative continuum Dyson-Schwinger approach, and solve a set of three coupled self-consistent integral equations for the fermion dressing functions. 
Our calculation contains three effects that have not previously been included:
\begin{enumerate}
\item vertex corrections constructed using an ansatz that preserves gauge invariance
\item magnetic effects (which correspond to contributions from transverse parts of the photon propagator)
\item full frequency dependence in dressing functions and loop integrals
\end{enumerate}
The BALL-CHIU calculation includes all three effects. The SHORT and COULOMB calculations use a modified vertex ansatz which is not fully gauge invariant, and neglect magnetic contributions. The COULOMB approximation differs form the SHORT calculation in that it also uses an expansion in $\delta\sim\{q_0,v_F\}$ (as discussed in Section \ref{nofreq-section}). 
Our results show that the BALL-CHIU and SHORT calculations give almost exactly the same value of the critical coupling. 
In QED$_{2+1}$ however, the Ball-Chiu vertex does effect the mass function \cite{fischer-2004}, and therefore it is somewhat surprising that there is no significant contribution to the result in a calculation using reduced QED$_{3+1}$ with a 2-brane.

All of our calculations use a perturbative Lindhard-type screening function in the photon propagator. A complete calculation would include self-consistently determined photon dressing functions. Since the photon polarization is determined from a fermion loop, and the fermion velocity renormalization is large, it is expected that a self-consistent calculation of photon screening could have a significant effect on the result. This calculation is currently in progress.

\leftline{\bf Acknowledgements}

This work was supported by the Natural Sciences and Engineering Research Council of Canada and by the Deutsche Forschungsgemeinschaft (DFG) under grant SM 70/3-1.

\end{document}